\documentclass[12pt,aps,onecolumn,superscriptaddress]{revtex4}
\usepackage{graphicx}
\usepackage{epsfig}
\usepackage{amssymb,amsmath}

\begin{document}        

\title{\large\bf\boldmath Isoscalar and isovector kaon form factors from 
$e^+e^-$ and $\tau$ data}

\begin{abstract}
The recent precise measurements of the $e^+e^-\to K_SK_L$ and 
$e^+e^-\to K^+K^-$ cross sections and the hadronic spectral function of 
the $\tau^-\to K^-K_S\nu_\tau$ decay are used to extract
the isoscalar and isovector electromagnetic kaon form factors and
their relative phase in a model independent way. The experimental
results are compared with a fit based on the vector-meson-dominance model.
\end{abstract}

\author{K.~I.~Beloborodov}
\email[e-mail: ]{K.I.Beloborodov@inp.nsk.su}
\affiliation{Budker Institute of Nuclear Physics, 630090 Novosibirsk, Russia}
\affiliation{Novosibirsk State University, 630090 Novosibirsk, Russia}

\author{V.~P.~Druzhinin}
\affiliation{Budker Institute of Nuclear Physics, 
630090 Novosibirsk, Russia}
\affiliation{Novosibirsk State University,
630090 Novosibirsk, Russia}

\author{S.~I.~Serednyakov}
\affiliation{Budker Institute of Nuclear Physics, 
630090 Novosibirsk, Russia}
\affiliation{Novosibirsk State University,
630090 Novosibirsk, Russia}

\maketitle

\section{Introduction}
Kaon electromagnetic form factors are the key objects in hadron physics
describing electromagnetic interaction of kaons and providing important
information about their internal structure.

In the timelike momentum-transfer region the form factors are
usually extracted from experimental data on the reactions
$e^+e^-\to K_SK_L$ and $e^+e^-\to K^+K^-$. In the resonance region
at center-of-mass (c.m.) energies $\sqrt{s}<2$ GeV, which we discuss
in this paper, a substantial improvement in the accuracy of 
these cross sections was achieved in the recent measurements in
the BABAR~\cite{BABAR-kckc,BABAR-kskl}, SND~\cite{SND-kckc}, and
CMD-3 experiments~\cite{CMD3-kskl,CMD3-kckc}. BABAR measured the
$e^+e^-\to K^+K^-$ and the $e^+e^-\to K_SK_L$ cross sections using the
initial-state-radiation method at the c.m. energies
$\sqrt{s}=0.98-4.85$ GeV and $\sqrt{s}=1.08-2.16$ GeV, respectively.
The SND and CMD-3 experiments used a direct scan. CMD-3 studied 
both the processes in the energy region near the $\phi$-meson peak, while
SND measured the $e^+e^-\to K^+K^-$ cross section in the range 
$\sqrt{s}=1.05-2.00$ GeV. New data are expected from the SND and CMD-3
experiments soon.

The $K^+K^-$ and $K_SK_L$ production Born cross sections are parametrized
in terms of the charged and neutral kaon form factors as follows
\begin{equation}
\sigma_{K^+K^-}(s)= \frac{\pi\alpha^2\beta^3}{3s}
\left|F_{K+}\right|^2C_{FS}(s),  \label{cskp}
\end{equation}
\begin{equation}
\sigma_{K_SK_L}(s)= \frac{\pi\alpha^2\beta^3}{3s}
\left|F_{K^0}\right|^2,  \label{csk0}
\end{equation}
where $\beta=\sqrt{1-4m_{K^{-(0)}}^2/s}$, and $m_{K^{-}}$ and $m_{K^{0}}$ are
the charged and neutral kaon masses for Eqs.~(\ref{cskp}) and (\ref{csk0}), 
respectively. The factor $C_{FS}$ is the final state correction (see, e.g., 
Ref.~\cite{fscor}). This correction has significant deviation from unity only
in a narrow interval near $K^+K^-$ threshold. 
The form factors $F_{K^+}$ and $F_{K^0}$ can be presented as a sum
of the isoscalar and isovector parts:
\begin{eqnarray}
F_{K^+}&=&F_{K^+}^{I=1}+F_{K^+}^{I=0},\\
F_{K^0}&=&F_{K^0}^{I=1}+F_{K^0}^{I=0}.
\end{eqnarray}
The isospin invariance gives following relations between amplitudes
for charged and neutral kaons~\cite{kuhn} 
\begin{eqnarray}
F_{K^0}^{I=0}& = &F_{K^+}^{I=0},\\
F_{K^0}^{I=1}& = &-F_{K^+}^{I=1}.
\end{eqnarray}
With this relations the cross sections proportional to squared moduli of the 
charged and neutral form factors can be expressed in term of isovector and 
isoscalar form factors $F_{K^+}^{I=0}$ and $F_{K^+}^{I=1}$
\begin{eqnarray}
|F_{K^+}|^2&=&|F_{K^+}^{I=1}|^2+2|F_{K^+}^{I=1}||F_{K^+}^{I=0}|cos(\Delta\phi_{K^+})+|F_{K^+}^{I=0}|^2,\\
|F_{K^0}|^2&=&|F_{K^+}^{I=1}|^2-2|F_{K^+}^{I=1}||F_{K^+}^{I=0}|cos(\Delta\phi_{K^+})+|F_{K^+}^{I=0}|^2,
\end{eqnarray}
where $\Delta\phi_{K^+}=\phi_{K^+}^{I=1}-\phi_{K^+}^{I=0}$ is a relative phase
between the isoscalar and isovector form factors. It is seen that
data on the $e^+e^-\to K_SK_L$ and $e^+e^-\to K^+K^-$ cross sections
do not allow to separate the isovector and isoscalar contributions in a 
model-independent way. However, additional experimental information can be
obtained from the $\tau^-\to K^-K^0\nu_\tau$ decay under the 
conserved-vector-current (CVC) hypothesis. Recently, the precision
measurement of the hadronic spectrum in this decay was performed
by the BABAR collaboration~\cite{BABAR-tau}.

The $\tau^-\to K^-K^0\nu_\tau$ differential decay rate as a function of
the $K^-K^0$ invariant mass $M$ normalized to the $\tau$ leptonic
width can be written as follows:
\begin{equation}
\frac{d{\cal B}(\tau^-\to K^-K^0\nu_\tau)}{{\cal B}(\tau^-\to\mu^-\bar{\nu_\mu}\nu_\tau)MdM}=
\frac{\left|V_{ud}\right|^2S_{EW}}{2m_\tau^2}
\left(1+\frac{2M^2}{m_\tau^2}\right)
\left(1-\frac{M^2}{m_\tau^2}\right)^2
\beta_-^3\left|F_{K^-K^0}(M)\right|^2,  \label{tau1}
\end{equation}
where 
$|V_{ud}|=0.97420\pm0.00021$~\cite{pdg} is the Cabibbo-Kobayashi-Maskawa
matrix element, $S_{EW}=1.0235\pm0.003$~\cite{davier} is the short-distance
electroweak correction, and $\beta_-=\sqrt{(1-(m_{K^{-}}+m_{K^{0}})^2/M^2)
(1-(m_{K^{-}}-m_{K^{0}})^2/M^2)}$. Here we introduce the form factor
$F_{K^-K^0}$. The CVC hypothesis in the limit of the isospin invariance
give the relation between this form factor and the isovector electromagnetic
form factor defined above~\cite{kuhn}
\begin{equation}
F_{K^-K^0} = - 2F_{K^+}^{I=1}. \label{tau2}
\end{equation}
It is tested for the $\tau^-\to \pi^-\pi^0\nu_\tau$ decay that the CVC 
hypothesis works with a few percent accuracy without introducing other
isospin-breaking corrections~\cite{davier1}.

Finally, using data on the $e^+e^-\to K_SK_L$ and $e^+e^-\to K^+K^-$ cross 
sections and the hadronic spectral function in the $\tau^-\to K^-K^0\nu_\tau$
decay we can separate the isoscalar and isovector contributions and determine
the moduli of the isoscalar and isovector form factors and the cosine of their
relative phase:
\begin{eqnarray}
|F_{K^+}^{I=1}|^2&=&4|F_{K^-K^0}|^2,\nonumber\\
|F_{K^+}^{I=0}|^2&=&\frac{|F_{K^+}|^2+|F_{K^0}|^2}{2}-|F_{K^+}^{I=1}|^2,\nonumber \\
cos(\Delta\phi_{K^+})&=&\frac{|F_{K^+}|^2-|F_{K^0}|^2}{2|F_{K^+}^{I=1}||F_{K^+}^{I=0}|}.\label{ffsep}
\end{eqnarray}
\begin{figure}
\setlength{\unitlength}{1mm}
\begin{picture}(70,50)
\put(0,0){
\epsfig{%
file=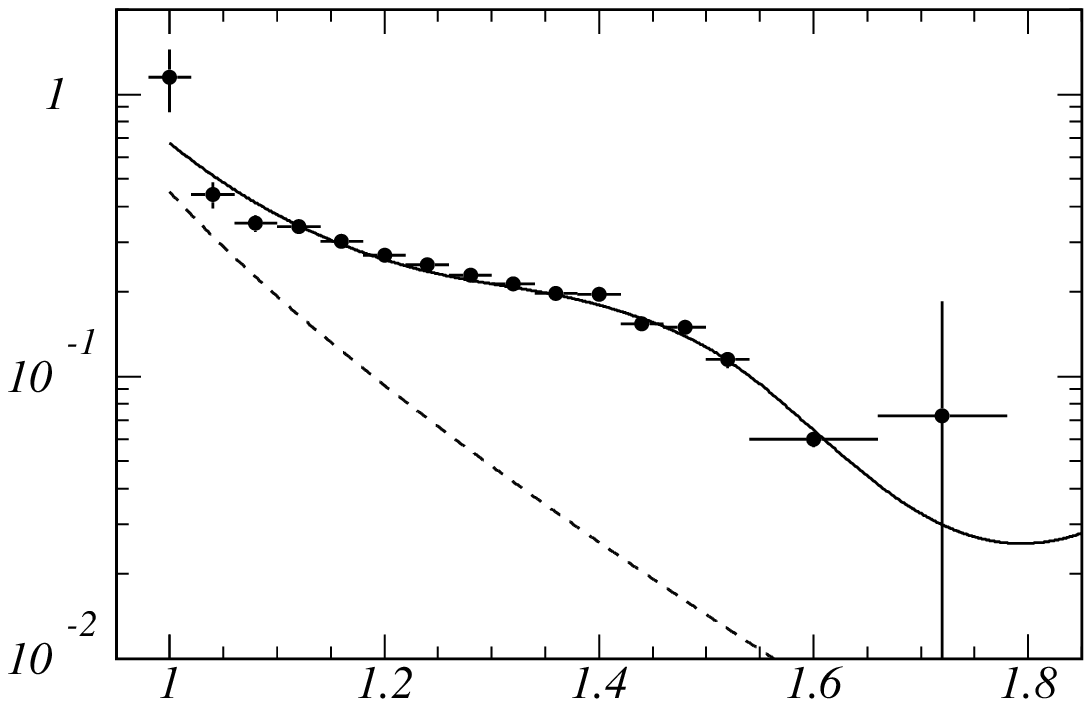,%
bbllx=0pt,%
bblly=0pt,%
bburx=340pt,%
bbury=227pt,%
width=70mm,%
height=50mm,%
clip=}
}
\put(52,-0.0){\small $\sqrt{s}$~,~GeV}
\put(-0.0,37){\rotatebox{90}{\small  $|F_{K^+}^{I=1}|^2$}}
\end{picture}
\begin{picture}(70,50)
\put(0,0){
\epsfig{%
file=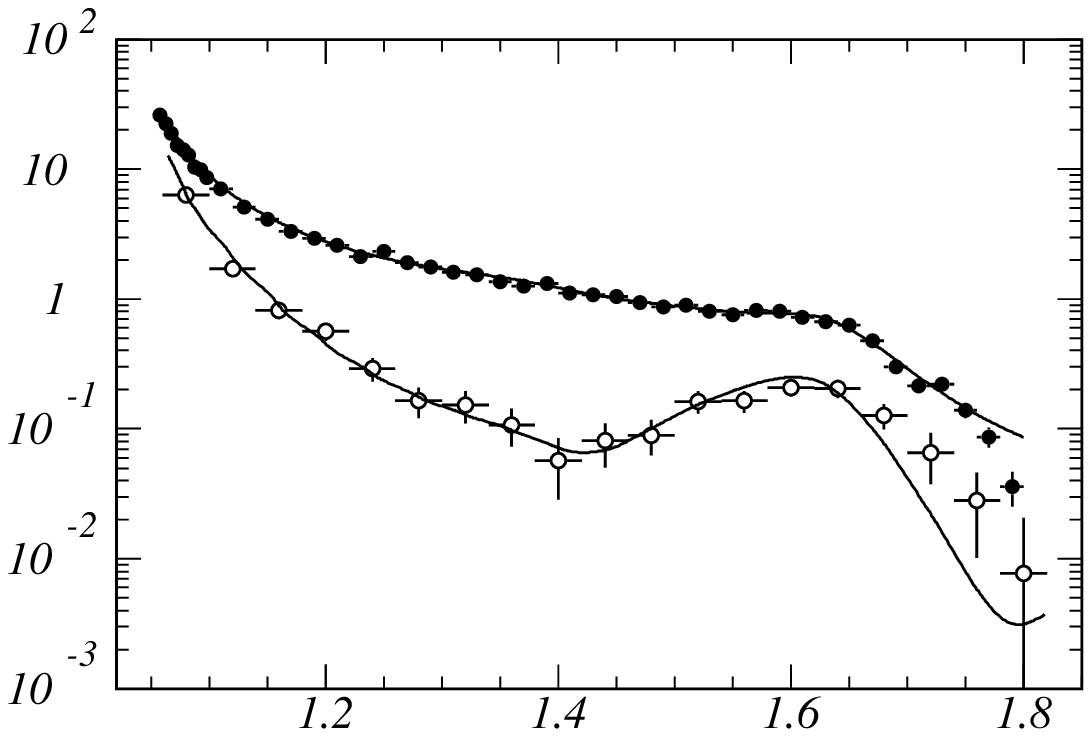,%
bbllx=0pt,%
bblly=0pt,%
bburx=340pt,%
bbury=227pt,%
width=70mm,%
height=50mm,%
clip=}
}
\put(52,-0.0){\small  $\sqrt{s}$~,~GeV}
\put(-0.0,27){\rotatebox{90}{\small  $|F_{K^+}|^{2}$,$|F_{K^o}|^{2}$}}
\end{picture}
\caption{\label{ffi1} Left panel: The isovector kaon form factor squared 
obtained from the $\tau^-\to K^-K^0\nu_\tau$ differential decay 
rate~\cite{BABAR-tau} as a function of $\sqrt{s}$. Right panel: The charged
(open circles) and neutral (filled circles) kaon form factors squared obtained
from the $e^+e^-\to K^+K^-$~\cite{BABAR-kckc} and 
$e^+e^-\to K_SK_L$~\cite{BABAR-kskl} cross section data respectively.
In the both panels, the solid curves represent the results of the fit 
(Model II) described in the text. The dashed curve in the left panel shows
the $\rho(770)$ contribution.}
\end{figure}

The isovector kaon form factor squared obtained using 
Eqs.~(\ref{tau1},\ref{tau2}) from the $\tau^-\to K^-K^0\nu_\tau$ differential 
decay rate~\cite{BABAR-tau} is shown in Fig.~\ref{ffi1}(left).  
The $\tau$ measurement covers the energy region from $m_{K^{-}}+m_{K^{0}}$
to $m_\tau$. This region is divided into two subregions, below and above
1.06 GeV, where the $\tau$ data should be treated in different way. Below
1.06 GeV the isoscalar form factor contains the resonance $\phi(1020)$,
which width is significantly smaller than the bin width in Fig.~\ref{ffi1}(left).  
Above 1.06 GeV excited vector resonances contributing to the form factors
have widths of about several hundred MeV. Therefore, we can use 
Eqs.~(\ref{ffsep}) to calculate the form factors in each energy bin of the
$\tau$ measurement without significant loss of information about their energy 
dependence. 

The charged and neutral kaon form factors above 1.06 GeV are shown
in Fig.~\ref{ffi1}(right). The neutral form factor is obtained using the
the most precise and extensive data on the $e^+e^-\to K_SK_L$ cross section
from the BABAR experiment~\cite{BABAR-kskl}.
The energy step in the $ K_SK_L$ and $\tau$ measurements is the same (40 MeV)
from 1.06 to 1.54 GeV. In the range 1.54-1.78 GeV corresponding the two
last wide bins of $\tau$ data, we average over 3 bins.
To obtain the charged form factor, the BABAR $e^+e^-\to K^+K^-$ data from 
Ref.~\cite{BABAR-kckc} are used. The SND measurement of the $e^+e^-\to K^+K^-$
cross section~\cite{SND-kckc} in the range 1.05-2.00 GeV having similar 
accuracy is in good agreement with the BABAR data.  It should be noted that 
the accuracy of the $e^+e^-\to K^+K^-$ cross section is significantly higher 
than those for the $ K_SK_L$ and $\tau$ measurements. In the energy region
of interest the energy step of the $K^+K^-$ measurement is 20 MeV. Therefore,
in further calculations the $K^+K^-$ data are averaged over 2 energy bins
in the energy range 1.06 to 1.54 GeV, and over 6 bins in the range from 
1.54 to 1.78 GeV, which corresponds the two last wide bins of $\tau$ data.
The the $ K_SK_L$ data in the latter range are averaged over 3 bins.
\begin{figure}
\setlength{\unitlength}{1mm}
\begin{picture}(70,50)
\put(0,0){
\epsfig{%
file=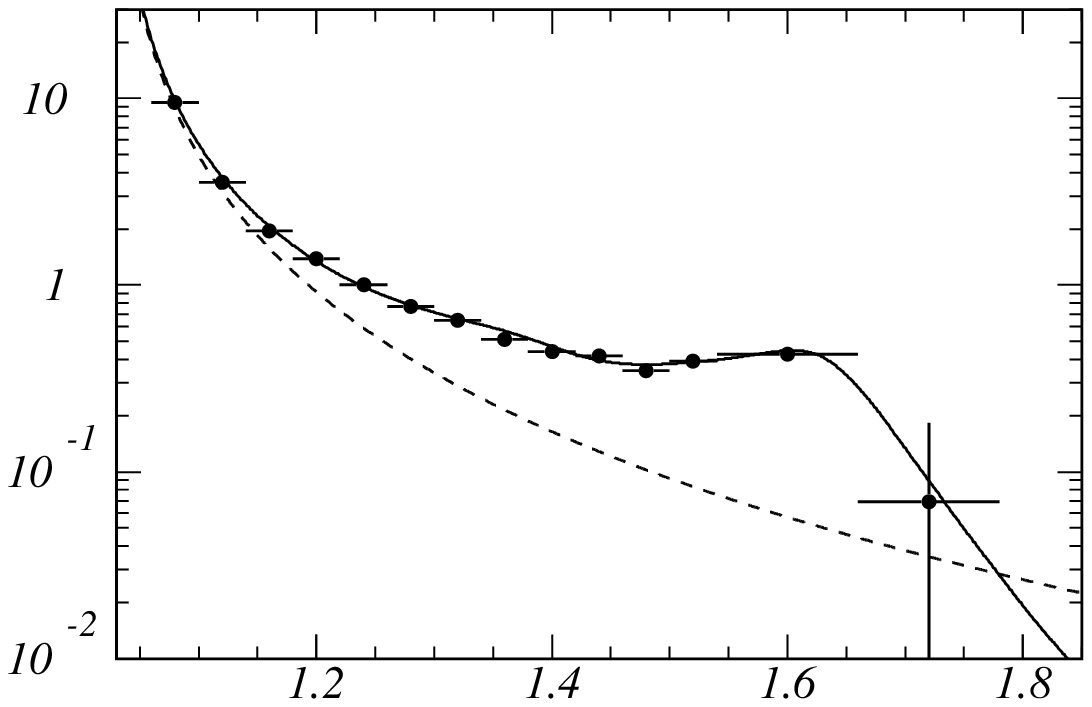,%
bbllx=0pt,%
bblly=0pt,%
bburx=340pt,%
bbury=227pt,%
width=70mm,%
height=50mm,%
clip=}
}
\put(52,-0.0){\small  $\sqrt{s}$~,~GeV}
\put(-0.0,37){\rotatebox{90}{\small  $|F_{K^+}^{I=0}|^{2}$}}
\end{picture}
\begin{picture}(70,50)
\put(0,0){
\epsfig{%
file=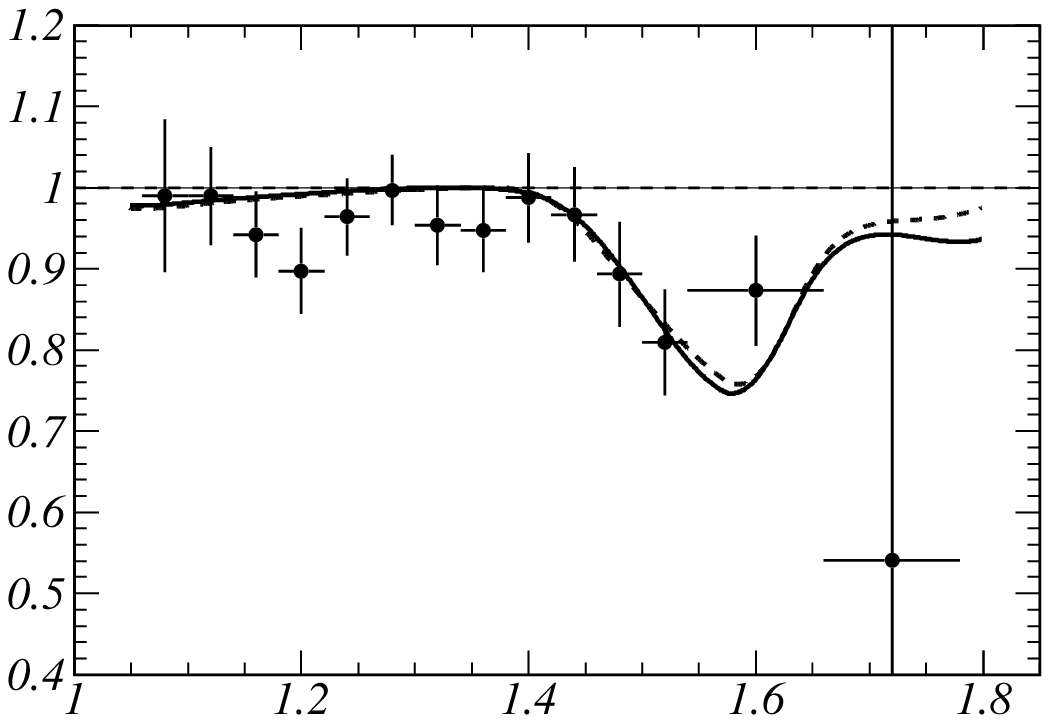,%
bbllx=0pt,%
bblly=0pt,%
bburx=340pt,%
bbury=227pt,%
width=70mm,%
height=50mm,%
clip=}
}
\put(52,-0.0){\small  $\sqrt{s}$~,~GeV}
\put(-0.0,31){\rotatebox{90}{\small  $cos(\Delta\phi_{K^+})$}}
\end{picture}
\caption{\label{ffc0} Left panel: The isoscalar kaon form factor squared
calculated using Eqs.~(\ref{ffsep}) from $e^+e^-$ and $\tau$ data as a function
of $\sqrt{s}$. The solid curve represents the results of the fit (Model II)
described in the text. The dashed curve shows the $\omega(782)$ and 
$\phi(1020)$ contribution. Right panel: The cosine of the relative phase 
between the isoscalar and isovector form factors calculated using 
Eqs.~(\ref{ffsep}) from $e^+e^-$ and $\tau$ data. The dashed and solid curves
represent the results of the fit with Model I and Model II described in the 
text, respectively.}
\end{figure}
The isoscalar kaon form and the cosine of the relative phase between
the isoscalar and isovector form factors calculated using Eqs.~(\ref{ffsep})
from $e^+e^-$ and $\tau$ data are shown in Fig.~\ref{ffc0}.

Both isoscalar and isovector form factors decrease monotonically in the
range below 1.4 GeV. This means that large contributions to the form
factors come from the tails of the $\rho(770)$ in the isovector
case, and $\omega(782)$ and $\phi(1020)$ in the isoscalar case. The latter
two contributions are expected to interfere constructively~\cite{kuhn}, making
the isoscalar form factor significantly larger than the isovector one. 
An unexpected feature of the form factors is the almost constant,
close-to-zero the phase difference between the isovector and isoscalar form 
factors in the energy range from 1.06 to 1.5 GeV. 
In this region, the resonances $\rho(1450)$ and $\omega(1420)$ are expected
to give contributions to the form factors, which interfere with the very 
different $\rho(770)$ isovector and $\omega(782)+\phi(1020)$ isoscalar 
amplitudes. Above 1.5 GeV, resonance structures related to the $\rho(1700)$,
$\omega(1650)$, and $\phi(1680)$ resonances are seen both in
the energy dependences of the form-factor moduli and the phase difference.

The second part of this article is devoted to the simultaneous fitting
of $e^+e^-$ and $\tau$ two-kaon data in the framework of the vector 
meson dominance (VMD) model assuming isospin invariance and CVC.
In this model, the amplitude of the single-photon transition
$A_{\gamma^*\to K\bar{K}}$ is described as a sum of amplitudes
of vector-meson resonances of the $\rho$, $\omega$, and $\phi$ families. 

The charged and neutral kaon cross sections are defined by the 
formulas~(\ref{cskp}) and (\ref{csk0}). For description 
of the charged and neutral form factors we use parametrization from
Ref.~\cite{kuhn}:
\begin{equation}
F_{K^+}(s)=\frac{1}{2}\sum_{V=\rho,\rho', {\ldots} }c_V BW_V+
\frac{1}{6}\sum_{V=\omega,\omega',{\ldots} }c_V BW_V+
\frac{1}{3}\sum_{V=\phi,\phi', {\ldots} }c_V BW_V,
\end{equation}
\begin{equation}
F_{K^0}(s)=-\frac{1}{2}\sum_{V=\rho,\rho', {\ldots} }c_V BW_V+
\frac{1}{6}\sum_{V=\omega,\omega',{\ldots} }c_V BW_V+
\frac{1}{3}\sum_{V=\phi,\phi', {\ldots} }c_V BW_V,
\end{equation}
where the sums are taken over the resonances of the $\rho$, $\omega$, or $\phi$
families, and the coefficients $c_V$ are real. We fit to the cross-section data
from the energy range below 2.1 GeV. The following resonances are included into
the fit: $\rho(770)$, $\rho(1450)$, $\rho(1700)$, and $\rho(2150)$ denoted as
$\rho$, $\rho'$, $\rho''$, and $\rho'''$, respectively, 
$\omega(782)$, $\omega(1420)$, $\omega(1680)$, and $\omega(2150)$ denoted as
$\omega$, $\omega'$, $\omega''$, and $\omega'''$, respectively,
$\phi(1020)$, $\phi(1680)$, and $\phi(2170)$ denoted as
$\phi$, $\phi'$, and $\phi''$, respectively. The $\rho'''$, $\omega'''$, and
 $\phi''$ are needed to describe the measured cross-section energy dependences
above 1.9 GeV. The partner of the $\rho(2150)$ resonance from the $\omega$ 
family is not observed yet. We introduce it into the fit with mass and
width equal to those for $\rho(2150)$.

The resonance line shapes are described by the Breit-Wigner function
\begin{equation}
BW_V(s) = \frac{M_V^2}{M_V^2-s-iM_V\Gamma_V(s)}, \label{BW}
\end{equation}
where $M_V$ and $\Gamma_V(s)$ are the resonance mass and energy dependent 
width. The widths for the $\omega$ and $\phi$-mesons take into 
account all significant decay modes: $\pi^+\pi^-\pi^0$, $\pi^0\gamma$, 
and $\pi^+\pi^-$ for $\omega$, and $K^+K^-$, $K_SK_L$, $\pi^+\pi^-\pi^0$, 
and $\eta\gamma$ for $\phi$. For the $\rho(770)$, we take into account 
the main $\pi^+\pi^-$ decay mode and the contribution of the
$\rho\to\omega\pi^0$ transition (see, for example, Ref.~\cite{ompi1}) 
with the coupling constant $g_{\rho\omega\pi}=15.9$ GeV$^{-1}$~\cite{ompi2}. 
For excited vector meson widths, only one dominant channel is used: $KK^*$ for
$\phi$-like resonances, $\omega\pi$ for $\rho'$, and $\rho\pi\pi$ for higher
excited $\rho$ states, $\rho\pi$ for $\omega'$, and $\omega\pi\pi$ for higher
excited $\omega$ states. The energy dependence of the partial widths are 
calculated using formulas from Refs.~\cite{achasov1,achasov2}. 

The $\tau^-\to K^-K^0\nu_\tau$ differential decay rate is described
by Eq.~(\ref{tau1}) with the form factor
\begin{equation}
F_{K^-K^0}(s)=-\sum_{V=\rho,\rho', {\ldots} }c_V BW_V.
\end{equation}

The data sets on the $e^+e^-\to K^+K^-$ and $e^+e^-\to K_SK_L$ cross
sections from CMD-3~\cite{CMD3-kskl,CMD3-kckc} in the $\phi$-meson region,
and from BABAR~\cite{BABAR-kckc,BABAR-kskl} in the 1.06-2.16 GeV region
are used in the fit. The BABAR $K^+K^-$ data below 1.06 GeV are not included
into the fit to avoid difficulties related to systematic difference in the
$\phi$-meson line shape and position between the CMD-3 and BABAR data sets.

The free fit parameters are the $\phi$-meson mass and width, a parameter
$\eta_\phi=g_{\phi K_S K_L}/g_{\phi K^+ K^-}$ describing the possible 
isospin-breaking difference between the $\phi\to K_S K_L$ and $\phi\to K^+K^-$
decay constants, and eight parameters $c_V$. The parameters $c_{\rho'''}$
and $c_{\phi''}$ are determined from the the conditions 
\begin{eqnarray}
\sum_{V=\rho,\rho', {\ldots} }c_V=1,\label{cVconstr1}\\
\frac{1}{3}\sum_{V=\omega,\omega',{\ldots} }c_V+\frac{2}{3}\sum_{V=\phi,\phi', {\ldots} }c_V=1,
\label{cVconstr2}
\end{eqnarray}
which provides the proper normalizations of the form factors $F_{K^+}(0)=1$
and $F_{K^0}(0)=0$. The parameter $c_{\omega'''}$ is taken to be equal
$c_{\rho'''}$, as it is expected from the quark model~\cite{kuhn}. 
The masses and widths of the $\rho$, $\omega$, and the excited vector 
resonances are fixed to their nominal values~\cite{pdg}. During the fit
they are allowed to vary within their uncertainties.
\begin{figure}
\setlength{\unitlength}{1mm}
\begin{picture}(70,50)
\put(0,0){
\epsfig{%
file=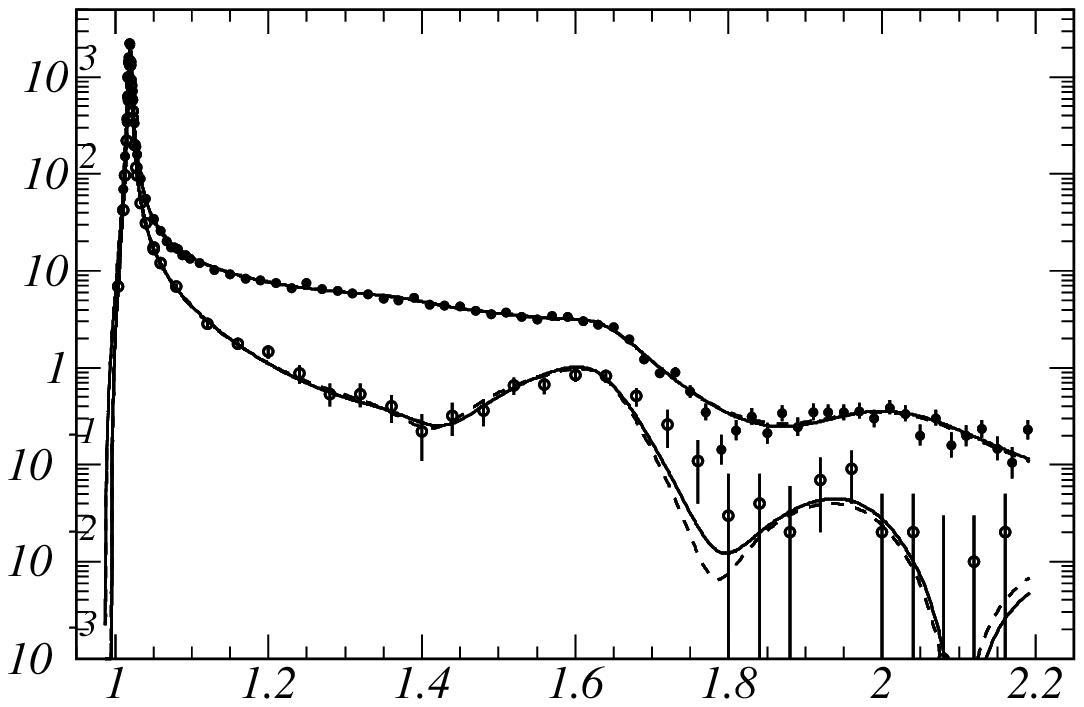,%
bbllx=0pt,%
bblly=0pt,%
bburx=340pt,%
bbury=227pt,%
width=70mm,%
height=50mm,%
clip=}
}
\put(52,-0.0){\small  $\sqrt{s}$~,~GeV}
\put(-0.0,37){\rotatebox{90}{\small  $\sigma$~,~nb}}
\end{picture}
\setlength{\unitlength}{1mm}
\begin{picture}(70,50)
\put(0,0){
\epsfig{%
file=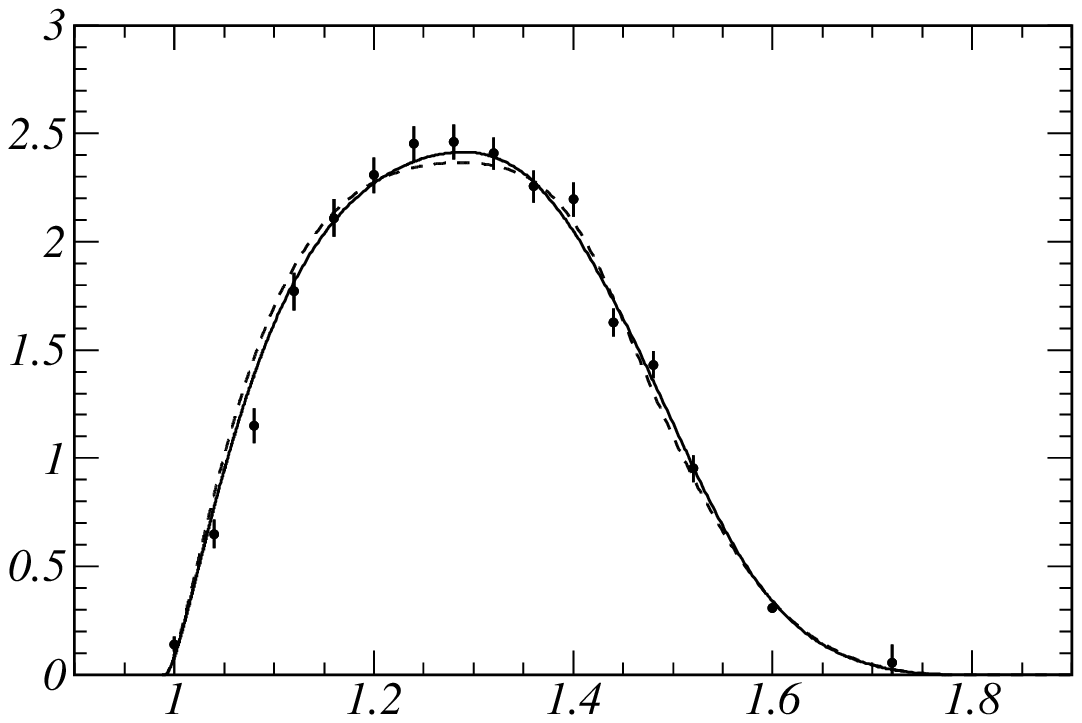,%
bbllx=0pt,%
bblly=0pt,%
bburx=340pt,%
bbury=227pt,%
width=70mm,%
height=50mm,%
clip=}
}
\put(52,-0.0){\small  $\sqrt{s}$~,~GeV}
\put(-0.0,26){\rotatebox{90}{\small  $1/B \cdot dB/dM$}}
\end{picture}
\caption{\label{css} Left panel: The $e^+e^-\to K^+K^-$ and 
$e^+e^-\to K_SK_L$ cross sections. Right panel: The $\tau^-\to K^-K^0\nu_\tau$
differential decay rate as a function of the $K^-K^0$ invariant mass.
The dashed and solid curves represent the results of the fit with Model I and
Model II described in the text, respectively.}
\end{figure}

The results of the fit are shown by the dashed curves (Model I) in 
Fig.~\ref{css} for the $e^+e^-\to K^+K^-$ and $e^+e^-\to K_SK_L$ cross sections
and the $\tau^-\to K^-K^0\nu_\tau$ differential decay rate, and in 
Fig.~\ref{ffc0}~(right) for the cosine of the relative phase between the 
isoscalar and isovector form factors. It is seen that the fitted curve does not
reproduce well the shape of the $\tau$-decay spectrum in 
Fig.~\ref{css}~(right). Therefore, we perform another fit (Model II),
in which the normalization constraints (\ref{cVconstr1}) and (\ref{cVconstr2})
are removed.  Due to closeness of the $\omega''$ and $\phi'$ masses the 
parameters $c_{\omega''}$ and $c_{\phi'}$ are strongly correlated and cannot
be determined in Model II independently. Therefore, the additional constraint 
$c_{\omega''}=c_{\rho''}$ is introduced.

The results of the fit with Model II are shown in Figs.~\ref{ffi1}, 
\ref{ffc0} and \ref{css} by the solid curves. This model describes the $\tau$
data significantly better and decreases the fit $\chi^2$ by 16 units. The 
resulting $\chi^2/\nu=183/142$, 
where $\nu$ is the number of degrees of freedom, is not quite good, but 
reasonable, taking into account that the systematic uncertainties of the 
measurements are not included into the fit. It should be also noted that the
sizable contribution to the $\chi^2$ (85 for 62 points) comes from the BABAR
$K^+K^-$ data, for which diagonal errors are used instead of the full error 
matrix. The sums on the left-hand sides of the normalization conditions
(\ref{cVconstr1}) and (\ref{cVconstr2}) are calculated to be $0.87\pm0.04$ 
and $0.98\pm0.05$, respectively. The 13\% deviation from unity for 
the first sum indicates that the the description of the $\rho$-like resonance
shapes, in particular the tail of the $\rho(770)$, in our fit model may
be not quite correct. The difference in the parameters $c_V$ between Model I
and Model II may be used as an estimate of their model uncertainty.

\begin{table}
\centering
\caption{\label{tab1}The fitted values of the coefficients $C_V$ in two models.
}
\begin{tabular*}{.8\textwidth}{l@{\extracolsep{\fill}}ccc}
\hline
V    & Model I & Model II   \\ 
\hline
$c_{\rho}$   & $1.162\pm0.005$  & $1.067\pm0.041$ \\
$c_{\rho'}$  & $-0.063\pm0.014$ & $-0.025\pm0.008$ \\
$c_{\rho''}$ & $-0.160\pm0.014$ & $-0.234\pm0.013$  \\
$c_{\rho'''}$& $\equiv 1-c_{\rho}-c_{\rho'}-c_{\rho''}$ & $0.063\pm0.007$ \\
\hline
$c_{\omega}$   & $1.26\pm0.06$   & $1.28\pm0.14$ \\
$c_{\omega'}$  & $-0.13\pm0.03$  & $-0.13\pm0.02$ \\
$c_{\omega''}$ & $-0.37\pm0.05$  & $\equiv c_{\rho''}$ \\
$c_{\omega'''}$& $\equiv c_{\rho'''}$ & $\equiv c_{\rho'''}$ \\
\hline
$c_{\phi}$   & $1.037\pm0.001$  & $1.038\pm0.001$ \\
$c_{\phi'}$  & $-0.117\pm0.020$ & $-0.150\pm0.009$ \\
$c_{\phi''}$ & $\equiv \frac{3}{2}-c_{\phi}-c_{\phi'}-\frac{1}{2}\sum_{V=\omega,\omega', {\ldots} }c_V$ & $ 0.089\pm0.015$  \\
\hline
$\chi^2/\nu$ & 199/143 & 183/142 \\
\hline
\end{tabular*}
\end{table}

The fitted value of the coefficient $\eta_\phi=0.990\pm0.001$ is found to be
consistent with unity. The $\eta_\phi$ value and fitted $\phi$-meson mass and
width, $M_{\phi}=1019.461\pm0.004$ and $\Gamma_\phi=4.248\pm0.006$ MeV,
agrees well with the values of these parameters obtained in 
Refs.~\cite{CMD3-kskl,CMD3-kckc}. The fitted values of the coefficients $C_V$
are listed in Table~\ref{tab1}. An interesting feature of the fits is a large 
deviation from the quark model predictions ($c_{\omega'}=c_{\rho'}$ and
$c_{\omega''}=c_{\rho''}$) for excited $\rho$ and $\omega$ resonances.
These deviations are needed, 
in particular, to provide the almost constant value of the phase difference
in the energy range 1.06--1.5 GeV, as it shown in Fig.~\ref{ffc0}(right).

We also perform a fit with an additional fit parameter $\alpha_{CVC}$ 
describing a possible deviation from the CVC hypothesis. This parameter is 
used as a scale factor to the $\tau$ data shown in Fig.~\ref{css}(right).
The fitted value of this parameter is $\alpha_{CVC}=0.986 (0.991)\pm0.020$ for 
Model I (II). This shows that the CVC hypothesis for the $K\bar{K}$ system 
works with a few percent accuracy.

In conclusion, we have used recent precise measurements of the $e^+e^-\to
K\bar{K}$ cross sections and the $K^-K_S$ spectrum in the $\tau^-\to 
K^-K_S\nu_\tau$ decay to separate the isoscalar and isovector electromagnetic
kaon form factors and determine the relative phase between them
in a model independent way. The latter shows an unexpected energy dependence
in the energy range from 1.06 to 1.5 GeV. It is almost constant and close to
zero. We have simultaneously fitted to the $e^+e^-\to K^+K^-$ and 
$e^+e^-\to K_SK_L$ cross-section data and the hadronic mass spectrum in the
$\tau^-\to K^-K_S\nu_\tau$ decay in the framework of the VMD model. The
fit reproduces data reasonably well and shows that the CVC hypothesis for the
$K\bar{K}$ system works with a few percent accuracy. To explain the specific
energy dependence of the relative phase between the isoscalar and isovector
form factors the large deviation from the quark model predictions 
for relations between the amplitudes of excited $\rho$ and $\omega$ 
resonances is required.

This work is supported in part by the RFBR grants 16-02-00327-a.

\end{document}